\documentclass[aps,pra,showpacs,twocolumn,groupedaddress]{revtex4}
\usepackage{graphicx,amsmath,amssymb}
\begin{document}
\newtheorem{Theorem}{Theorem}
\newtheorem{Lemma}{Lemma}

\title{Quantum adiabatic evolutions that can't be used to design efficient algorithms}

\author{Zhaohui Wei}
\email{weich03@mails.tsinghua.edu.cn }
\author{Mingsheng Ying}
\email{yingmsh@mail.tsinghua.edu.cn }

\affiliation{ State Key Laboratory of Intelligent Technology and
Systems, Department of Computer Science and Technology, Tsinghua
University, Beijing, China, 100084}

\begin{abstract}

Quantum adiabatic computation is a novel paradigm for the design of
quantum algorithms, which is usually used to find the minimum of a
classical function. In this paper, we show that if the initial
hamiltonian of a quantum adiabatic evolution with a interpolation
path is too simple, the minimal gap between the ground state and the
first excited state of this quantum adiabatic evolution is an
inverse exponential distance. Thus quantum adiabatic evolutions of
this kind can't be used to design efficient quantum algorithms.
Similarly, we show that a quantum adiabatic evolution with a simple
final hamiltonian also has a long running time, which suggests that
some functions can't be minimized efficiently by any quantum
adiabatic evolution with a interpolation path.

\end{abstract}

\maketitle

Quantum computation has attracted a great deal of attention in
recent years, because some quantum algorithms show that the
principles of quantum mechanics can be used to greatly enhance the
efficiency of computation. Recently, a new novel quantum computation
paradigm based on quantum adiabatic evolution has been proposed
\cite{FGGS00}. We call quantum algorithms of this paradigm quantum
adiabatic algorithms. In a quantum adiabatic algorithm, the
evolution of the quantum register is governed by a hamiltonian that
varies continuously and slowly. At the beginning, the state of the
system is the ground state of the initial hamiltonian. If we encode
the solution of the algorithm in the ground state of the final
hamiltonian and if the hamiltonian of the system evolves slowly
enough, the quantum adiabatic theorem guarantees that the final
state of the system will differ from the ground state of the final
hamiltonian by a negligible amount. Thus after the quantum adiabatic
evolution we can get the solution with high probability by measuring
the final state. For example, Grover's algorithm has been
implemented by quantum adiabatic evolution in \cite{RC02}. Recently,
the new paradigm for quantum computation has been tried to solve
some other interesting and important problems
\cite{RAO03,DKK02,TH03,TDK01,FGG01}.

Usually, except in some simple cases, a decisive mathematical
analysis of a quantum adiabatic algorithm is not possible, and
frequently even the estimation of the running time is very
difficult. Sometimes we have to conjecture the performance of
quantum adiabatic algorithms by numerical simulations, for example
in \cite{FGG01}. In this paper, we estimate the running time of a
big class of quantum adiabatic evolutions. This class of quantum
adiabatic evolutions have a simple initial hamiltonian and a
universal final hamiltonian. We show that the running time of this
class of quantum adiabatic evolutions is exponential of the size of
problems. Thus they can't be used to design efficient quantum
algorithms. We noted that E. Farhi et al. have got the similar
result by a continuous-time version of the BBBV oracular proof
\cite{BBBV97} in \cite{FGGN05}. However, our proof is based on the
quantum adiabatic theorem, which is much simpler and more direct.
Furthermore, our result can be generalized from the case of linear
path to the case of interpolation paths. Besides, by the symmetry of
our proof it is easy to prove that a quantum adiabatic evolution
that has a simple final hamiltonian and a universal final
hamiltonian also has a long running time, which can be used to
estimate the worst performance of some quantum adiabatic algorithms.

For convenience of the readers, we briefly recall the local
adiabatic algorithm. Suppose the state of a quantum system is
$|\psi(t)\rangle(0\leq t\leq T)$, which evolves according to the
Schr\"{o}dinger equation
\begin{equation}
i\frac{d}{dt}|\psi(t)\rangle=H(t)|\psi(t)\rangle,
\end{equation}
where $H(t)$ is the Hamiltonian of the system. Suppose $H_0=H(0)$
and $H_1=H(T)$ are the initial and the final Hamiltonians of the
system. Then we let the hamiltonian of the system vary from $H_0$ to
$H_1$ slowly along some path. For example, a interpolation path is
one choice,
\begin{equation}
H(t)=f(t)H_0+g(t)H_1,
\end{equation}
where $f(t)$ and $g(t)$ are continuous functions with $f(0)=g(T)=1 $
and $f(T)=g(0)=0$ ($T$ is the running time of the evolution). Let
$|E_0,t\rangle$ and $|E_1,t\rangle$ be the ground state and the
first excited state of the Hamiltonian at time t, and let $E_0(t)$
and $E_1(t)$ be the corresponding eigenvalues. The adiabatic theorem
\cite{LIS55} shows that we have
\begin{equation}
|\langle E_0,T|\psi(T)\rangle|^{2}\geq1-\varepsilon^2,
\end{equation}
provided that
\begin{equation}
\frac{D_{max}}{g_{min}^2}\leq\varepsilon,\ \ \ \ 0<\varepsilon\ll1,
\end{equation}
where $g_{min}$ is the minimum gap between $E_0(t)$ and $E_1(t)$
\begin{equation} g_{min}=\min_{0\leq t \leq T}[E_1(t)-E_0(t)],
\end{equation}
and $D_{max}$ is a measurement of the evolving rate of the
Hamiltonian
\begin{equation}
D_{max}=\max_{0\leq t \leq
T}|\langle\frac{dH}{dt}\rangle_{1,0}|=\max_{0\leq t \leq T}|\langle
E_1,t|\frac{dH}{dt}|E_0,t\rangle|.
\end{equation}

Before representing the main result, we give the following lemma.

\begin{Lemma}Suppose $f:\{0,1\}^n\rightarrow R$ is a function that is bounded
by a polynomial of n. Let $H_0$ and $H_1$ be the initial and the
final hamiltonians of a quantum adiabatic evolution with a linear
path $H(t)$. Concretely,
\begin{equation}
H_0=I-|\alpha\rangle\langle\alpha|,
\end{equation}
\begin{equation}
H_1=\sum\limits_{z=1}^{N}{f(z)|z\rangle\langle z|},
\end{equation}
\begin{equation}
H(t)=(1-t/T)H_0+(t/T)H_1,
\end{equation}
where, $T$ is the running time of the quantum adiabatic evolution
and
\begin{equation}
|\alpha\rangle=|\hat{0^n}\rangle=\frac{1}{\sqrt{N}}\sum\limits_{i=1}^{N}{|i\rangle},
\ \ N=2^n.
\end{equation}
Then we have
\begin{equation}
g_{min} < \frac{2}{2^{n/2-n/100}}.
\end{equation}
Thus $T$ is exponential in $n$.
\end{Lemma}

{\it Proof.} Let
$$H(s)=(1-s)(I-|\alpha\rangle\langle\alpha|)+s\sum\limits_{z=1}^{N}{f(z)|z\rangle\langle z|},$$
where $s=t/T$. Suppose $\{f(z),1 \leq z\leq N\}=\{a_{i},1\leq i\leq
N\}$ and $a_1\leq a_2 \leq ... \leq a_N$. Without loss of
generality, we suppose $a_1=0$. Otherwise we can let
\begin{equation}
H(s)=H(s)-I\times\min_{1\leq i \leq N}a_i,
\end{equation}
which doesn't change $g_{min}$ of $H(s)$. We also suppose
$a_i<a_{i+1},1<i<N-1$(Later we will find that this restriction can
be removed).

Now we consider $A(\lambda)$, the characteristic polynomial of
$H(s)$. It can be proved that
\begin{align}
A(\lambda)=&\prod_{i=1}^N{(1-s-\lambda+sa_i)} \cr &
-\frac{1-s}{N}\sum_{j=1}^N{\prod_{k\neq j}^D{(1-s-\lambda+sa_k)}}.
\end{align}

For every $s\in(0,1)$, we have $A(0)>0$ and $A(1-s)<0$. Because
$A(\lambda)$ is a polynomial, $A(\lambda)$ has a root $\lambda_1(s)$
in the interval $(0,1-s)$. Similarly, in each of the intervals
$(1-s+sa_k,1-s+sa_{k+1})(1\leq k\leq N-1)$ there is a root
$\lambda_{k+1}(s)$. It can be proved that in the interval
$(0,\lambda_1(s))$, $A(\lambda)>0$. Otherwise if for some
$\lambda_0\in(0,\lambda_1(s))$, $A(\lambda_0)<0$, there will be
another root in the interval $(0,\lambda_0)$. In this case the
number of the eigenvalues of $H(s)$ is more than $N$, which is a
contradiction. Similarly we have $A(\lambda)<0$ for interval
$(\lambda_1(s),\lambda_2(s))$ and we have $A(\lambda)>0$ for
interval $(\lambda_2(s),1-s+sa_2)$.

\begin{figure}[htb]
\centerline{\includegraphics[angle=0,width=3.3in]{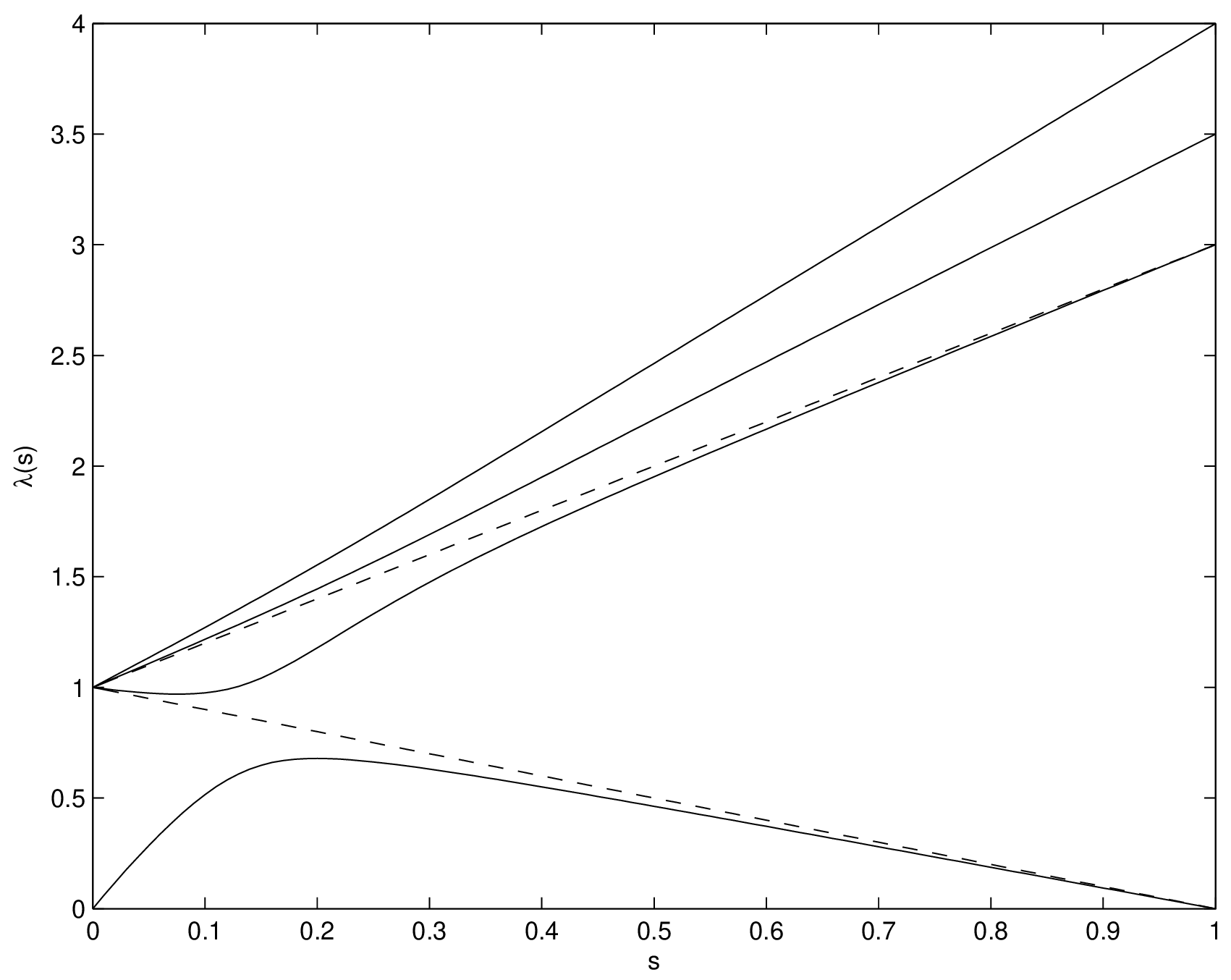}}
\caption{The two dashed lines are $\lambda(s)=1-s$ and
$\lambda(s)=1-s+sa_2$, and the solid lines are the four lowest
eigenvalue curves of $H(s)$, where $N=16$, $a_1=0$, and $a_i=2+i/2$
for $1<i\leq 16$.} \label{fig:gaps_Id}
\end{figure}

Consider a line $\lambda_2^{'}(s)=1-(1-1/m)s$ in the
$s$-$\lambda(s)$ plane, where $m=poly(n)$ or a positive polynomial
in n. Suppose we can find a $m$ that $a_2>1/m$ for every $n$ big
enough. Then we know that for every big $n$, the line
$\lambda_2^{'}(s)$ lies in the region between lines $\lambda=1-s$
and $\lambda=1-s+sa_2$. By solving the inequation $A(1-(1-1/m)s)<0$
we can get which part of the line $\lambda_2^{'}(s)$ lies in the
region between lines $\lambda=1-s$ and the eigenvalue curve
$\lambda_2(s)$. The result is, when $s\in(0,s_2)$ the line
$\lambda_2^{'}(s)$ lies above $\lambda_2(s)$ and when $s\in(s_2,1)$
the eigenvalue curve $\lambda_2(s)$ lies above $\lambda_2^{'}(s)$,
where
\begin{equation}
s_2=\frac{1}{1+\frac{N}{\sum\limits_{j=1}^{N}{\frac{1}{a_j-\frac{1}{m}}}}}.
\end{equation}
Similarly, we consider another line $\lambda_1^{'}(s)=1-(1+1/m)s$.
By similar analysis, we get that when $s\in(0,s_1)$ the line
$\lambda_1^{'}(s)$ lies above $\lambda_1(s)$ and when $s\in(s_1,1)$
the eigenvalue curve $\lambda_1(s)$ lies above $\lambda_1^{'}(s)$,
where
\begin{equation}
s_1=\frac{1}{1+\frac{N}{\sum\limits_{j=1}^{N}{\frac{1}{a_j+\frac{1}{m}}}}}.
\end{equation}
It can be proved that for any fixed positive polynomial $m$
\begin{equation}
s_{1}<s_{2}
\end{equation}
if $n$ is big enough. Now we consider the interval $(s_1,s_2)$. In
this interval, the eigenvalues curves $\lambda_1(s)$ and
$\lambda_2(s)$ all lies between the lines $\lambda_1^{'}(s)$ and
$\lambda_2^{'}(s)$. At the same time, it is easy to know that the
gap between the lines $\lambda_1^{'}(s)$ and $\lambda_2^{'}(s)$ is
less than $2/m$. Thus the minimal gap between $\lambda_1(s)$ and
$\lambda_2(s)$ is also less than $2/m$. That is to say,
\begin{equation}
g_{min}<2/m.
\end{equation}
Obviously, to get Eq.(17) the restriction $a_2>1/m$ above can be
removed, because if $a_2<1/m$, the gap between $\lambda_1(s)$ and
$\lambda_2(s)$ is less than $1/m$ when $s$ is near 1, then we also
have Eq.(17). Furthermore, the restriction $a_i<a_{i+1},1<i<N-1$ can
also be removed. If $a_i=a_{i+1}$ for some $i$, we can give a very
small disturbance to $H_1$, which make every $a_i$ different, while
$g_{min}$ doesn't change too much (for example, we can let the
change of $g_{min}$ much less than $1/m$).

Similarly, Supposing $m = 2^{n/2-n/100}$, we also have $s_{1}<s_{2}$
if $n$ is big enough. At this time, for any $s\in(s_1, s_2)$ the gap
between $\lambda_1(s)$ and $\lambda_2(s)$ is less than
$\frac{2}{2^{n/2-n/100}}$. So we have
\begin{equation}
g_{min}<\frac{2}{2^{n/2-n/100}}.
\end{equation}
In fact, 100 in Eq.(18) can be replaced by any big natural number.
By the quantum adiabatic theorem, the running time of this quantum
adiabatic evolution is exponential in $n$. That completes the proof
of this lemma. \hfill $\Box$

Lemma 1 shows that, to find the minimum of the function $f(x)$
effectively using the quantum adiabatic algorithms, the initial
hamiltonian $H_0$ can't be too simple (See also \cite{FGGN05}). If
we set the initial hamiltonian according to the structure of the
function $f(x)$, the effect maybe better. For example in section 7.1
of \cite{DMV01},
\begin{equation}
H_0=\sum\limits_{z\in\{0,1\}^n}{w(z)|\hat{z}\rangle\langle
\hat{z}|},
\end{equation}
and
\begin{equation}
H_1=\sum\limits_{z\in\{0,1\}^n}{w(z)|z\rangle\langle z|},
\end{equation}
where $H_0$ is diagonal in the Hadamard basis with the bit values
\begin{equation}
|\hat{0}\rangle = \frac{1}{\sqrt{2}}(|0\rangle+|1\rangle), \  \
|\hat{1}\rangle = \frac{1}{\sqrt{2}}(|0\rangle-|1\rangle),
\end{equation}
and $w(z)=z_1+z_2+...+z_n$. In this quantum adiabatic evolution, the
initial hamiltonian reflect the structure of the function that we
want to minimize. $g_{min}$ of this evolution is independent of $n$,
and the quantum algorithm consisted by this evolution is efficient.

Noted that Lemma 1 shows that the time complexity of the quantum
adiabatic algorithm for the hidden subgroup problem proposed in
\cite{RAO03} is exponential in the number of input qubits
\cite{RAO06}. Similarly, the main result of \cite{ZH06} can also be
got again via Lemma 1, which was also pointed out in \cite{FGGN05}.

In Lemma 1, the path of quantum adiabatic evolutions is linear. The
following theorem shows that this can be generalized.

\begin{Theorem}Suppose $H_0$ and $H_1$ given by Eq. (7) and Eq. (8)
are the initial and the final hamiltonians of a quantum adiabatic
evolution. Suppose this quantum adiabatic evolution has a
interpolation path
\begin{equation}
H(t)=f(t)H_0+g(t)H_1.
\end{equation}
Here $f(t)$ and $g(t)$ are arbitrary continuous functions, subject
to the boundary conditions
\begin{equation}
f(0) = 1 \ \ \ g(0) = 0,
\end{equation}
\begin{equation}
f(T) = 0 \ \ \ g(T) = 1,
\end{equation}
and
\begin{equation}
c_1<f(t) + g(t)< c_2, \ \ 0\leq t \leq T,
\end{equation}
where, $T$ is the running time of the adiabatic evolution and $c_1$
and $c_2$ are positive real numbers. Then we have
\begin{equation}
g_{min} < \frac{2c_2}{2^{n/2-n/100}}.
\end{equation}
Thus $T$ is exponential in $n$.
\end{Theorem}

{\it Proof.} Note that
\begin{equation}
H(t)=(f(t)+g(t))(\frac{f(t)}{f(t)+g(t)}H_0+\frac{g(t)}{f(t)+g(t)})H_1,
\end{equation}
and $\frac{f(t)}{f(t)+g(t)}$ is a continuous functions whose range
of function is $[0,1]$. Suppose the gap of the ground state and the
first excited state of the quantum adiabatic evolution
$H'(t)=(1-t/T)H_0+(t/T)H_1$ arrives at its minimum at $t_0
\in[0,T]$, then the corresponding gap of $H(t)$ at $t'_0$ will be
less than $\frac{2c_2}{2^{n/2-n/100}}$, where
$\frac{g(t'_0)}{f(t'_0)+g(t'_0)}=t_0/T$.

That completes the proof of this Theorem. \hfill $\Box$

We have shown that a simple initial hamiltonian is bad for a quantum
adiabatic evolution. Similarly, a simple final hamiltonian is also
bad. First we represent the following lemma.

\begin{Lemma}Suppose $f:\{0,1\}^n\rightarrow R$ is a function that is bounded
by a polynomial of n. Let $H_0$ and $H_1$ are the initial and the
final hamiltonians of a quantum adiabatic evolution with a linear
path $H(t)$. Concretely,
\begin{equation}
H_0=\sum\limits_{z=1}^{N}{f(z)|\hat{z}\rangle\langle\hat{z}|},
\end{equation}
\begin{equation}
H_1=I-|x\rangle\langle x|, \ \ 1\leq x\leq N,
\end{equation}
\begin{equation}
H(t)=(1-t/T)H_0+(t/T)H_1,
\end{equation}
where, $H_0$ is diagonal in the Hadamard basis and $T$ is the
running time of the quantum adiabatic evolution. Then we have
\begin{equation}
g_{min} < \frac{2}{2^{n/2-n/100}}.
\end{equation}
Thus $T$ is exponential in $n$.
\end{Lemma}

{\it Proof.} Let $$H'(s)=(1-s)H_1+sH_0,$$ and
$$H''(s)=(H^{\bigotimes n})H'(s)(H^{\bigotimes n}),$$
where $s=t/T$ and $H$ is the Hadamard gate. First, by symmetry it's
not difficult to prove that $H'(s)$ and $H(s)$ have the same
$g_{min}$. Second, $H'(s)$ and $H''(s)$ have the same characteristic
polynomial, then they also have the same $g_{min}$. So the $g_{min}$
of $H''(s)$ is the minimal gap that we want to estimate. On the
other hand, it also can be proved that $H''(s)$ has the same
characteristic polynomial as Eq.(13) no matter what $x$ is. Thus
according to Lemma 1 we can finish the proof. \hfill $\Box$

Analogously, Lemma 2 can also be generalized to the case of
interpolation paths.

\begin{Theorem}Suppose $H_0$ and $H_1$ given by Eq. (28) and Eq. (29)
are the initial and the final hamiltonians of a quantum adiabatic
evolution. Suppose this quantum adiabatic evolution has a
interpolation path
\begin{equation}
H(t)=f(t)H_0+g(t)H_1.
\end{equation}
Here $f(t)$ and $g(t)$ are arbitrary continuous functions, subject
to the boundary conditions
\begin{equation}
f(0) = 1 \ \ \ g(0) = 0,
\end{equation}
\begin{equation}
f(T) = 0 \ \ \ g(T) = 1,
\end{equation}
and
\begin{equation}
c_1<f(t) + g(t)< c_2, \ \ 0\leq t \leq T,
\end{equation}
where, $T$ is the running time of the adiabatic evolution and $c_1$
and $c_2$ are positive real numbers. Then we have
\begin{equation}
g_{min} < \frac{2c_2}{2^{n/2-n/100}}.
\end{equation}
Thus $T$ is exponential in $n$.
\end{Theorem}

If $f(z)$ arrives at its minimum when $z=z_0$ and if for any $z\neq
z_0$ $f(z)$ has the same value, Eq.(8) will have a form similar
Eq.(29). Theorem 2 shows that if we use a quantum adiabatic
evolution with a interpolation path to minimize a function of this
kind, the running time will be exponential in $n$ no matter what
$H_0$ is. For example, in quantum search problem $f(z)$ is of this
form. Again we show that quantum computation can't provide
exponential speedup for search problems. Furthermore, Theorem 2 can
help us with other problems. For some quantum adiabatic algorithms,
we may use it to consider the possible worst case. If in some case
$f(z)$ only has two possible values and arrives at the minimum at
only one point, we can say that the worst case performance of the
quantum adiabatic algorithm with a interpolation path that minimizes
$f(z)$ is exponential.

In conclusion, we have shown in a quantum adiabatic algorithm if the
initial hamiltonian or the final hamiltonian is too simple, the
performance of the algorithm will be very bad. Thus, we have known
that when designing quantum algorithms, some quantum adiabatic
evolutions are hopeless. Furthermore, we also know that for some
function $f(z)$, we can't use any quantum adiabatic algorithm with a
interpolation path to minimize it effectively.

\end{document}